\documentclass[11pt,twoside]{article}
\usepackage{asp2014}
\usepackage{color}
\usepackage{ulem}

\aspSuppressVolSlug
\resetcounters

\newcommand{\gr}{$\gamma$-ray}
\newcommand{\dml}{dm-$\lambda$}
\newcommand{\jgr}{J. Geophys. Res.}

\begin{document}

\title{Science with an ngVLA: Radio Observations of Solar Flares}
\author{D. E. Gary,$^1$ T. S. Bastian,$^2$, B. Chen,$^1$, G. D. Fleishman,$^1$ and L. Glesener$^3$}
\affil{$^1$New Jersey Institute of Technology, Newark, NJ, USA; \email{dgary@njit.edu, bin.chen@njit.edu, gfleishm@njit.edu}}
\affil{$^2$National Radio Astronomy Observatory, Charlottesville, VA, USA; \email{tbastian@nrao.edu}}
\affil{$^3$University of Minnesota, Minneapolis, MN, USA; \email{glesener@umn.edu}}

\paperauthor{D. E. Gary}{dgary@njit.edu}{0000-0003-2520-8396}{New Jersey Institute of Technology}{}{Newark}{NJ}{07102}{USA}
\paperauthor{T. S. Bastian}{tbastian@nrao.edu}{0000-0002-0713-0604}{National Radio Astronomy Observatory}{}{Charlottesville}{VA}{22903}{USA}
\paperauthor{B. Chen}{bin.chen@njit.edu}{0000-0002-0660-3350}{New Jersey Institute of Technology}{}{Newark}{NJ}{07102}{USA}
\paperauthor{G. D. Fleishman}{gfleishm@njit.edu}{}{New Jersey Institute of Technology}{}{Newark}{NJ}{07102}{USA}
\paperauthor{L. Glesener}{glesener@umn.edu}{}{New Jersey Institute of Technology}{}{Newark}{NJ}{07102}{USA}

\begin{abstract}
Solar flares are due to the catastrophic release of magnetic energy in the Sun's corona, resulting in plasma heating, mass motions, particle acceleration, and radiation emitted from radio to $\gamma$-ray wavelengths. They are associated with global coronal eruptions of plasma into the interplanetary medium---coronal mass ejections---that can result in a variety of ``space weather'' phenomena. Flares release energy over a vast range of energies, from $\sim\!10^{23}$ ergs (nanoflares) to more than $10^{32}$ ergs. Solar flares are a phenomenon of general astrophysical interest, allowing detailed study of magnetic energy release, eruptive processes, shock formation and propagation, particle acceleration and transport, and radiative processes. Observations at radio wavelengths offer unique diagnostics of the physics of flares. To fully exploit these diagnostics requires the means of performing  time-resolved imaging spectropolarimetry. Recent observations with the Jansky Very Large Array (JVLA) and the Expanded Owens Valley Solar Array (EOVSA), supported by extensive development in forward modeling, have demonstrated the power of the approach. The ngVLA has the potential to bring our understanding of flare processes to a new level through its combination of high spatial resolution, broad frequency range, and imaging dynamic range---especially when used in concert with multi-wavelength observations and data at hard X-ray energies.
\end{abstract}

\section{Introduction}
Solar flares are produced by the catastrophic release of magnetic energy in the Sun's corona. The sporadic explosive release of excess (i.e. non-potential) magnetic energy is a ubiquitous and fundamental property of magnetized astrophysical plasmas, which is generally accompanied by shocks, turbulence, and acceleration of charged particles. Such energy releases may be driven by a supernova explosion or other gamma-ray burst progenitor. Acceleration of charged particles in solar flares and their transport share common mechanisms with the acceleration and propagation of the cosmic rays produced in such powerful explosions. Other cases may be powered by gas accretion onto a massive central object such as a black hole. In particular, this includes Galactic microquasars/X-ray binaries produced by accretion from a normal star onto a compact stellar mass object (black hole or neutron star) or Active Galactic Nuclei driven by a similar yet even more powerful phenomenon. The accretion forms a disk whose interaction with the central object produces, via unknown mechanisms, a single or pair of oppositely directed highly collimated relativistic jets. These jets produce most of the observed radio emission and contribute essentially to emission at other wavelength ranges, implying that the jets are the main or substantial source of the accelerated particles.  Potentially similar jet phenomena are seen and can be studied in great detail in the solar case \citep[e.g.][]{2018arXiv180600858G}.

Solar flares result in plasma heating, the acceleration of electrons and ions to high energies, mass motions, shocks, and electromagnetic radiation across the entire spectrum, from radio to \gr\ wavelengths. Flares are often associated with global eruptions, coronal mass ejections (CMEs), that can eject $\gtrsim\!10^{15}$ gm of material at speeds up to 2000 km s$^{-1}$, as well as smaller-scale collimated jets.  Jets, flares, and CMEs can drive interplanetary disturbances and produce electromagnetic and particle radiation that can have a direct and adverse effects on Earth and near-Earth environment (``space weather''). Hence, while solar flares are of considerable astrophysical interest as an energetic phenomenon in their own right, flares and CMEs are of practical interest because of their potential for affecting modern technological infrastructure and astronaut safety. In this chapter, we touch on just two outstanding problems---magnetic energy release in flares and  particle acceleration---and discuss the potential role of the ngVLA for addressing them. The range of solar and space physics science topics that can be addressed by the ngVLA is much broader, however \citep[cf.][]{bastian_ngvla,fleishman_ngvla}.

\section{Background and Motivation}

Solar flares are efficient particle accelerators, converting  several tenths of the available free energy stored in stressed magnetic fields to high-energy particles.  Electrons can be promptly accelerated to energies of 10s of MeV and ions to energies $\sim$1 GeV/nuc \citep{1997JGR...10214631M}. Energetic particles on the Sun emit radiation at radio, X-ray, \gr\ wavelengths. Extreme ultraviolet (EUV) and soft X-ray (SXR) emission mainly arise from thermal plasma heated to $\sim\!20$ MK. Hard X-ray (HXR) emission typically arises from nonthermal bremsstrahlung from electrons with energies of $\sim$10~keV to a few $\times$ 100~keV interacting with dense chromospheric material or coronal plasma. Continuum \gr\ emission ($E >$ few $\times$ 100 keV) arises from electron-proton and electron-electron bremsstrahlung and pion decay, and gamma-ray lines are caused by electron-positron annihilation, neutron capture, and nuclear de-excitation. Thermal plasma produces radio waves via bremsstrahlung and gyroresonance emission. Suprathermal electrons (few keV to a few 10s of keV) may produce intense coherent radio emission and energetic electrons (10s of keV to many MeV) emit nonthermal gyrosynchrotron (GS) radio radiation. Together, HXR, \gr, and radio observations provide a powerful and complementary suite of perspectives from which to study magnetic energy release and particle acceleration, addressing questions such as:

\begin{itemize}
\item Where does magnetic energy release occur?
\item Where are electrons accelerated? By what mechanisms?
\item What is the electron energy distribution function and what is its spatiotemporal evolution?
\item What are the dominant transport effects for the energetic electrons?
\end{itemize}

With their sensitivity to non-equilibrium particle distributions and magnetic fields, radio observations offer a number of unique diagnostics of flare physics. A revolutionary advance in radio instrumentation is the ability to map the Sun at radio wavelengths, with high spectral and temporal resolution over a broad range of frequencies. The combination of spatially resolved imaging and high spectral resolution is called imaging spectroscopy or, when circular polarization is added, imaging spectropolarimetry. When coupled with high time resolution, {\sl dynamic} imaging spectropolarimetry becomes possible, bringing with it the ability to image every frequency-time bin in the broadband dynamic spectrum.

Broadband imaging spectroscopy with high temporal resolution (0.01--1~s) is key for resolving the dynamic behavior of solar radio bursts.  To fully cover the range of phenomena of interest to solar physicists, a very broad bandwidth is needed.  For example, the main microwave component of solar flares is the inherently broadband mechanism of gyrosynchrotron (GS) radiation, and the spectrum is further broadened by non-uniformities in magnetic field strength, density, and particle energy in the flaring region.  The typical GS spectrum peaks in the range 5--10~GHz \citep{1975SoPh...44..155G}, but the peak can move to much higher frequencies \citep[$>35$~GHz in some cases; see, e.g.,][]{1986A&A...157...11K, 2004SoPh..223..181R, 2018ApJ...856..111L}, or to lower frequencies \citep[$<1$~GHz; e.g.,][]{ 1994SoPh..152..409L, Fl_etal_2016}, reflecting the wide ranges of magnetic field strength (few~G to few~kG) and particle energy (few~keV to few~MeV) that are possible. In addition, coherent bursts occur in a large fraction of solar flares, and although a discrete burst may be intrinsically narrow band, multitudes of features can appear simultaneously across a wide range of frequency, or rapidly drift over many octaves of frequency.

Two existing radio instruments are pioneering the exploitation of these new capabilities: the {\sl Jansky Very Large Array} (JVLA) and the {\sl Expanded Owens Valley Solar Array} (EOVSA).  We draw on examples of observations from these instruments here. The next generation Very Large Array (ngVLA), if designed to take solar observing requirements into account, can expand greatly on these capabilities by providing improved frequency coverage and dynamic range. A cursory discussion of solar science with the Square Kilometer Array (SKA) was done by \citet{2004NewAR..48.1511B}, but in its current design state its importance for solar observations has not been established.

\section{Probing Magnetic Energy Release with Coherent Radio Bursts}

Coherent radio emission occurs at resonant frequencies of the coronal medium via a coherent radiation process: the electron plasma frequency $\nu_{pe}$, the electron gyrofrequency $\Omega_{Be}$, or combinations and harmonics thereof---e.g., the upper hybrid frequency. Coherent radio bursts are frequently observed in solar flares of all sizes.  They can dominate the radio spectrum at decimeter and meter wavelengths \citep{1998ARA&A..36..131B, 2004ApJ...605..528N}, and are sometimes seen at MW frequencies as high as $\sim9$~GHz.

There are two main radiation mechanisms for solar coherent radio emission at decimetric wavelengths (\dml\ hereafter). One of them is plasma radiation, which is due to the nonlinear conversion of plasma Langmuir waves to transverse radio waves. It occurs at the fundamental and/or second harmonic of the electron plasma frequency $\nu_{pe}=\sqrt{n_e e^2/\pi m_e}\sim n_{10}^{1/2}$ GHz, providing a direct measurement of the local electron density in the radio source $n_{10}$ (units $10^{10}$ cm$^{-3}$). The other is electron cyclotron maser emission (ECME), which arises around the lowest harmonics ($s=$~1, 2, 3) of the electron gyrofrequency yet the source has to be reasonably tenuous to let the radiation to escape. Bursts due to ECME can provide valuable information about the magnetic field in the corona. Dm-$\lambda$ coherent radio bursts correspond to plasma density and magnetic field characteristic of the low solar corona, where the primary energy release and particle acceleration in solar flares are presumed to occur \citep{1998ARA&A..36..131B, 2011SSRv..159..225W}. Therefore, observations of \dml\ coherent bursts provide the means of tracing the presence and motion of electrons accelerated in and around the energy release site. Most previous studies have depended on single dish radio dynamic spectral observations of \dml\ bursts \citep[][and references therein]{2017RvMPP...1....5M}. In the absence of spatial information, a definitive association of these coherent radio bursts to the larger magnetic environment, needed to establish their role in particle acceleration, is lacking. Dynamic imaging spectroscopy surmounts this limitation.

\begin{figure}\centering
\includegraphics[width=1.0\textwidth, clip]{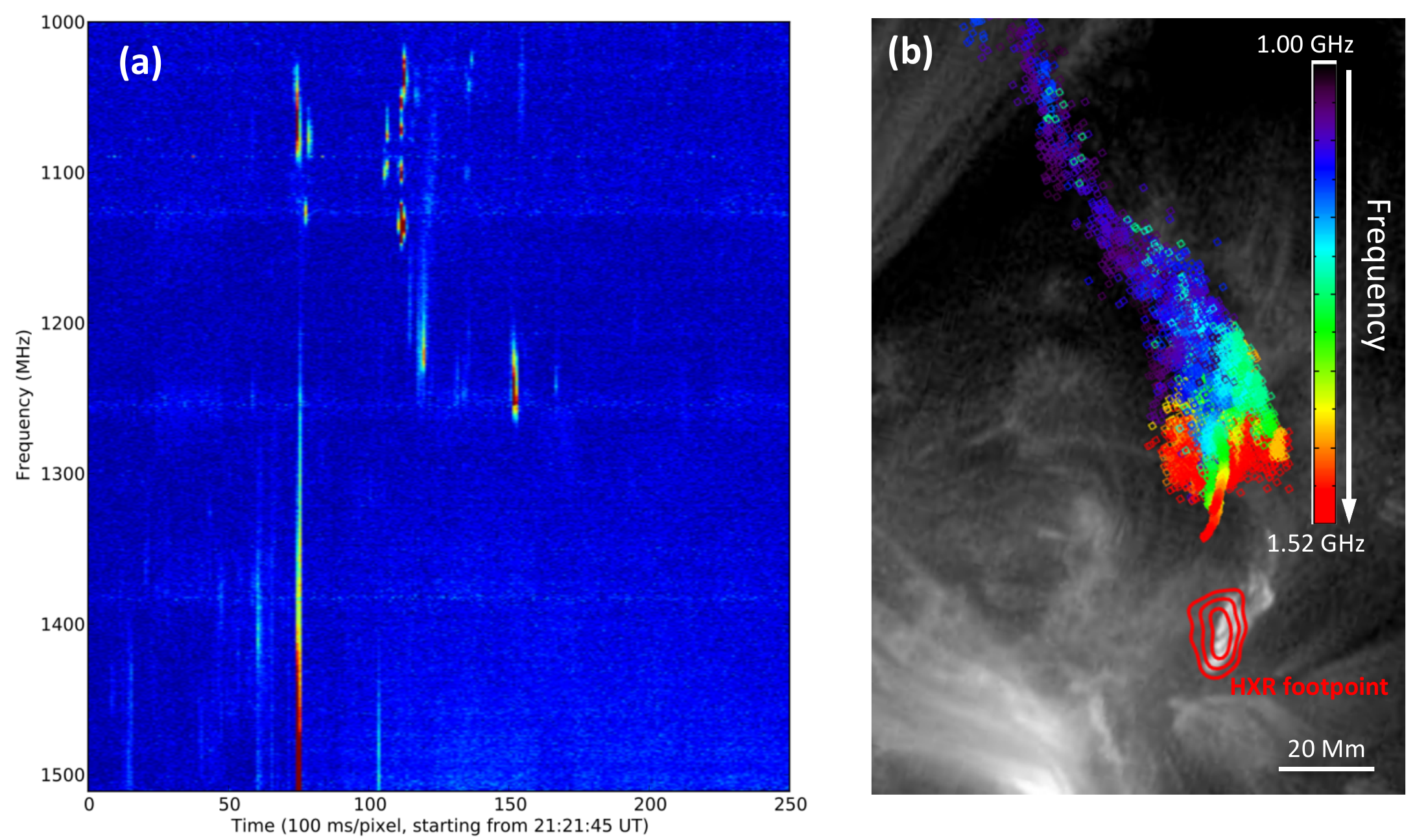}
\caption{\label{tp3}
(a) Type III radio bursts observed by the JVLA associated with a solar jet. The bursts are visible in the radio dynamic spectrum (radio intensity as a function of time and frequency, colored from blue to red in increasing intensity) as bright, nearly vertical short-lived features. (b) Emission centroids of \dml\ bursts shown in (a), colored from blue to red in increasing frequency. The ensemble of the centroids show electron beam trajectories in projection. Red contours are HXR emission due to downward-moving nonthermal electrons colliding with the dense solar chromosphere. The background EUV image from SDO/AIA shows coronal loops and the jet event. (Adapted from \citealt{2013ApJ...763L..21C}.)
 }
\end{figure}

An excellent example is \dml\ type III radio bursts (Figure~\ref{tp3}). These bursts are the signature of flare-accelerated, fast electron beams ($\sim$0.1--0.5$c$) propagating along newly-reconnected magnetic field lines. An electron beam propagating upward in the corona encounters plasma with a decreasing density, thus generating radio waves at progressively lower frequencies. In a dynamic spectrum, a type IIIdm burst is seen as a fast-drift feature with a negative frequency slope ($d\nu/dt<0$). Similarly, a beam propagating downward will result in a fast-drift burst in the dynamic spectrum with a positive slope ($d\nu/dt>0$). As an electron beam propagates along newly reconnected magnetic field lines, its trajectory traces the instantaneous magnetic topology of the on-going reconnection, which is usually invisible at other wavelengths unless the loops have been pre-heated or populated with dense plasma. Interestingly, sometimes both the positive- and negative-drift type III bursts are seen simultaneously. These burst pairs are typically interpreted as oppositely-directed fast electron beams that emerge from a common region, presumably the energy release site. These radio bursts have the potential to pinpoint the particle acceleration site, diagnose its physical properties \citep{1997ApJ...480..825A,2014RAA....14..773R}, and allow the wider magnetic field connectivity to be mapped.

\begin{figure}\centering
\includegraphics[width=1.0\textwidth, clip]{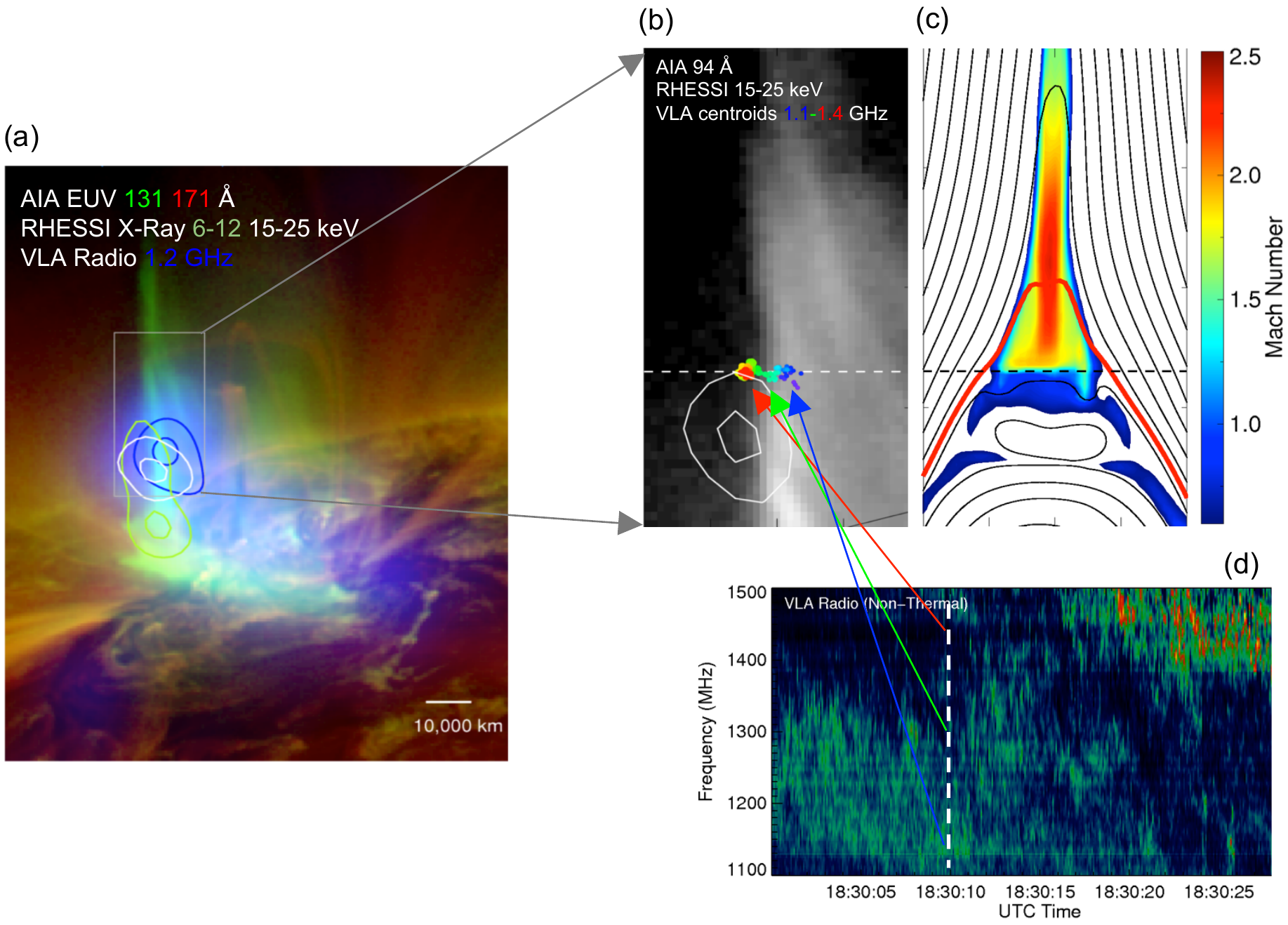}
\caption{\label{spikes}
Radio observation interpreted as a termination shock in an eruptive solar flare. (a) Composite image of an eruptive solar flare in radio (JVLA; blue), EUV (SDO/AIA, green and red), and HXR (\textit{Reuven Ramaty High Energy Solar Spectroscopic Imager} (RHESSI), white contours). (b) A dynamic surface-like feature revealed by the spatial distribution of radio source centroids in different frequencies. It is interpreted as the front of a fast-mode termination shock formed at the top of flaring arcades seen in EUV and HXR. Each colored dot corresponds to the centroid location of a spike burst in the radio dynamic spectrum at a given time and frequency (panel d). This shock is well-reproduced by magnetohydrodynamics simulations (panel c) as a sharp transition layer through which super-magnetosonic reconnection outflows become sub-magnetosonic when they encounter the underlying, obscuring dense loops. (Adapted from \citealt{2015Sci...350.1238C}.)
 }
\end{figure}

Dynamic structures such as shocks, magnetic islands, and turbulence are also believed to exist in or around the magnetic energy release region. Some, if not all, are thought to play an important role in the the particle acceleration process  \citep{1986ApJ...305..553F, 1996ApJ...461..445M, 2006Natur.443..553D}. Certain coherent radio burst types, such as \dml\ spike bursts (Figure~\ref{spikes}d) and drifting pulsating structures, have been suggested to arise from interactions between accelerated electrons and dynamic structures in or around the energy release region \citep{1985SoPh...96..357B,Fl_Meln_1998,2000A&A...360..715K,Rozh_etal_2008}. The location and dynamics of such radio bursts, obtained by radio dynamic imaging spectroscopy, provide unique access to these dynamic structures. This is demonstrated in a recent study by \citet{2015Sci...350.1238C} based on JVLA observations of an eruptive solar flare, in which dynamic spectral imaging of \dml\ spike bursts was utilized to map the radio source as a function of frequency and time (Figure~\ref{spikes}b). The spatial distribution of the burst centroids outlines a dynamic surface at the top of the flare arcades, interpreted by \citet{2015Sci...350.1238C} as the front of a fast-mode termination shock (Figure~\ref{spikes}c), which likely contributed to accelerating electrons to nonthermal energies. Given that individual spikes are short-lived (milliseconds to tens of ms) and narrowband (at the order of 1\%), the spatiotemporal distribution of some important physical parameters in the source region, e.g., plasma density or magnetic field as well as their level of fluctuations (possibly due to turbulence), may be derived or inferred from the data with high accuracy, provided that the responsible emission mechanism is identified.

The ngVLA will be capable of imaging these coherent radio bursts over the 1.2-3.5~GHz band with high spectral resolution (0.1--1\%) and time resolution (10 ms), with excellent dynamic range. One of the limitations of the current VLA is its relatively low dynamic range, which in the presence of bright coherent emissions renders the weaker incoherent emission undetectable.  Combined with simultaneous observations in the higher frequency bands \citep[cf.,][]{spikes}, the ngVLA will be able to link coherent \dml\ emissions to incoherent GS emission from much higher energy electrons, for which quantitative diagnostics are available for parameters in the flaring region, as described in the following section.

\section{Particle Acceleration and Transport}

Radio emission processes (e.g. \citealt{1985ARA&A..23..169D}; \citealt{1998ARA&A..36..131B}; \citealt{2004ASSL..314...71G};
\citealt{2013ASSL..388.....F}) are known to depend strongly on the magnetic field strength in the source region due to two effects: (1) the Lorentz force causes the emitting electrons to spiral in the magnetic field to produce gyroemission, which takes the form of gyroresonance (GR) emission for thermal electrons in the hot corona of active regions, and nonthermal GS emission for mildly-relativistic electrons in flaring magnetic loops; (2) the plasma itself is birefringent, with differing index of refraction (e.g. \citealt{2013ASSL..388.....F}) for the ordinary (o) and extraordinary (x) magnetoionic modes, which leads to a magnetic-field-dependent polarization of each type of emission, including free-free (FF) emission \citep{2004ASSL..314..115G}. All three emission mechanisms, GR, GS, and FF, have been exploited in countless studies of radio emission from the Sun to constrain properties of the thermal plasma, the distribution function of energetic electrons, and the magnetic field strength in the source.  However, such estimates based on either spatial or spectral information alone generally suffer from a large number of ambiguities that require varying degrees of assumptions, or else require special conditions where the assumptions and ambiguities can be minimized. These ambiguities can be largely eliminated when dynamic imaging spectropolarimetry is available.

\begin{figure}\centering
\includegraphics[width=0.8\textwidth, clip]{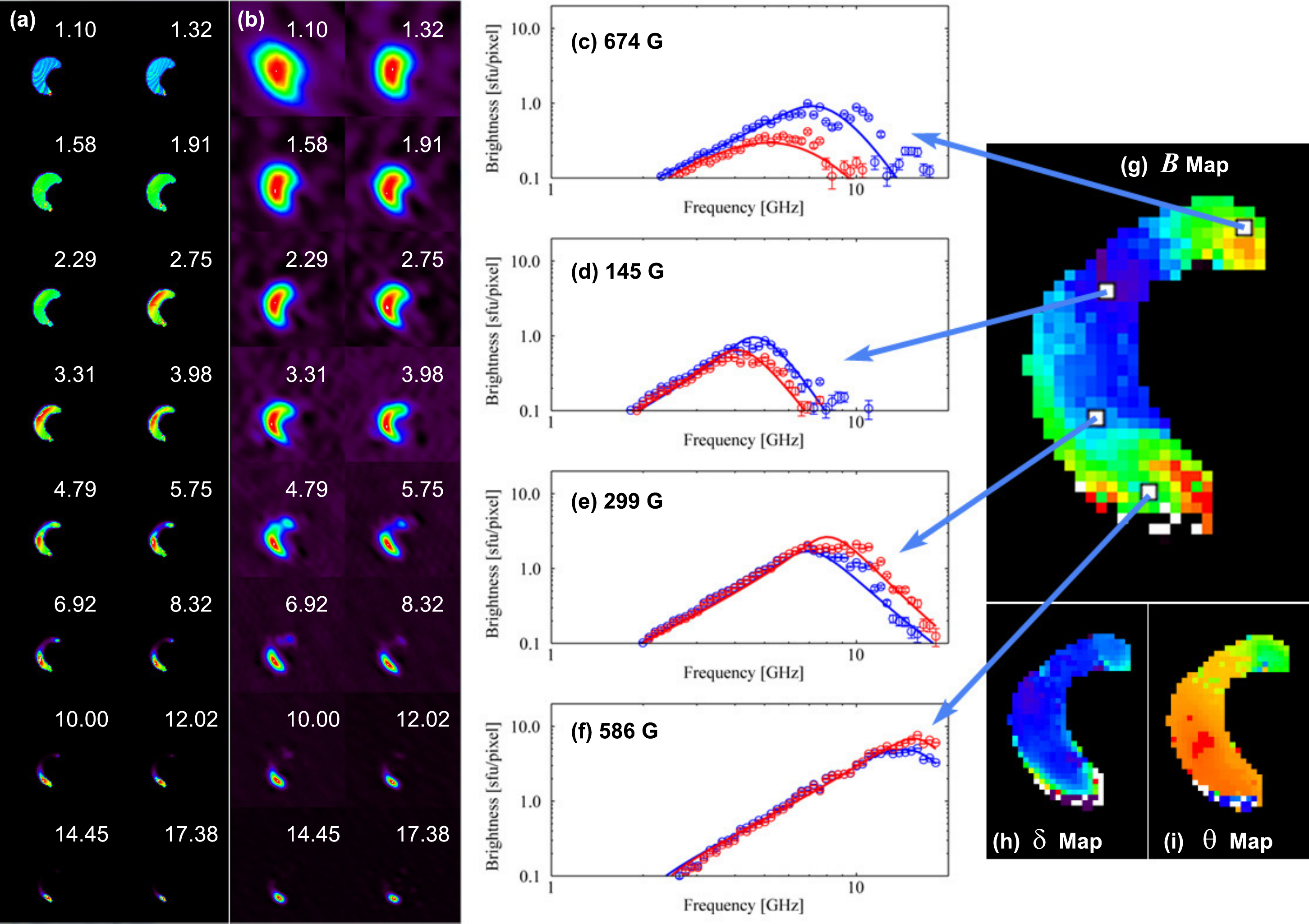}
\caption{\label{fig1}
(a) Model images of a flaring loop at 16 frequencies in the range 1-18 GHz. (b) Reconstructed images after sampling with the EOVSA snapshot array coverage. Note the transformation from a uniform source at 1.2 GHz to a loop-top dominated source at 2.75 GHz to a footpoint dominated source above 12 GHz. (c-f) Polarized brightness (intensity) spectra at 64 frequencies from the reconstructed images (points with error bars), overlaid with red (RCP) and blue (LCP) 6-parameter fits. (g) Pixel-by-pixel magnetic field map derived from spectral fits, with the four specific locations shown that correspond to the plots in c-f. (h-i) Maps of power-law index (constant $\delta = 5$ was used), and line-of-sight angle to B ($64\deg < \theta < 123\deg$ along loop), derived from spectral fits. (Adapted from \citealt{2013SoPh..288..549G}.)
 }
\end{figure}

Nonthermal GS emission is of particular interest. The sensitivity of its emissivity to the magnetic field means that it is visible wherever nonthermal electrons are present and have access to coronal magnetic loops.
Figure~\ref{fig1} shows an illustrative example. This model representation of a snapshot (1-s) data cube has both spatial and spectral information in both senses of circular polarization, which provides spatially-resolved, polarized brightness-temperature spectra at each point in the field of view.

The model is obtained by placing a realistic distribution of nonthermal electrons and background thermal plasma in a magnetic loop (derived from an extrapolation of photospheric magnetic field measurements) and calculating the resulting emission including both GS and FF contributions. The calculation of the microwave emission is performed using the best currently available Fast Gyrosynchrotron Codes \citep{2010ApJ...721.1127F}, which include both accurate treatment of the GS and FF processes in each voxel along the given line of sight and the solution of the radiation transfer along the line of sight, taking into account optical-depth and Razin suppression effects as well as frequency-dependent mode coupling. These model images are then sampled with EOVSA's array configuration, with suitable thermal noise added, to obtain the images in Fig. 1b.  By stacking the multi-frequency images into a data cube, spatially-resolved spectra at every point in the flaring region are obtained in right and left circular polarization (RCP and LCP), shown with red and blue symbols, respectively, in Fig. 1c-f.  Multi-parameter fits to the spectra (solid red and blue curves) are performed as described by \citet{2009ApJ...698L.183F}. One of the parameters, magnetic field strength $B$, is indicated in each panel.

When such fits are done at every pixel, parameter maps such as those shown in panels Fig.~3g-3i can be constructed. Fig.~3g shows a distribution of the magnetic field strength in the loop and Fig.~3i shows the distribution of the magnetic field angle to the line of sight. Fig.~3h shows the spectral index $\delta$ of the power-law electron distribution used for the simulation. In the more general case, the {\sl spatiotemporal} evolution of the electron distribution function will be inferred from the brightness temperature spectra. These measurements will place {\sl new} and powerful constraints on the mechanisms responsible for electron acceleration (shocks, turbulence, electric fields) and their subsequent transport (trapping, scattering, precipitation).

Fig.~3 is a illustrative simulation. Some recent EOVSA observations of the X8.3 solar flare of 2017-Sep-10 \citep[see][]{2018ApJ...863...83G} are shown in Fig.~4. The partially occulted flare, a classic flux-rope eruption seen in profile above the solar limb, was well observed by other instruments.  Microwave imaging spectroscopy provides an entirely new addition to the study of such events, by providing the spatial distribution of otherwise invisible high-energy electrons as well as direct measurement of plasma parameters, including magnetic field, as a function of position and time in the radio-emitting volume.

\begin{figure}\centering
\includegraphics[width=1.0\textwidth, clip]{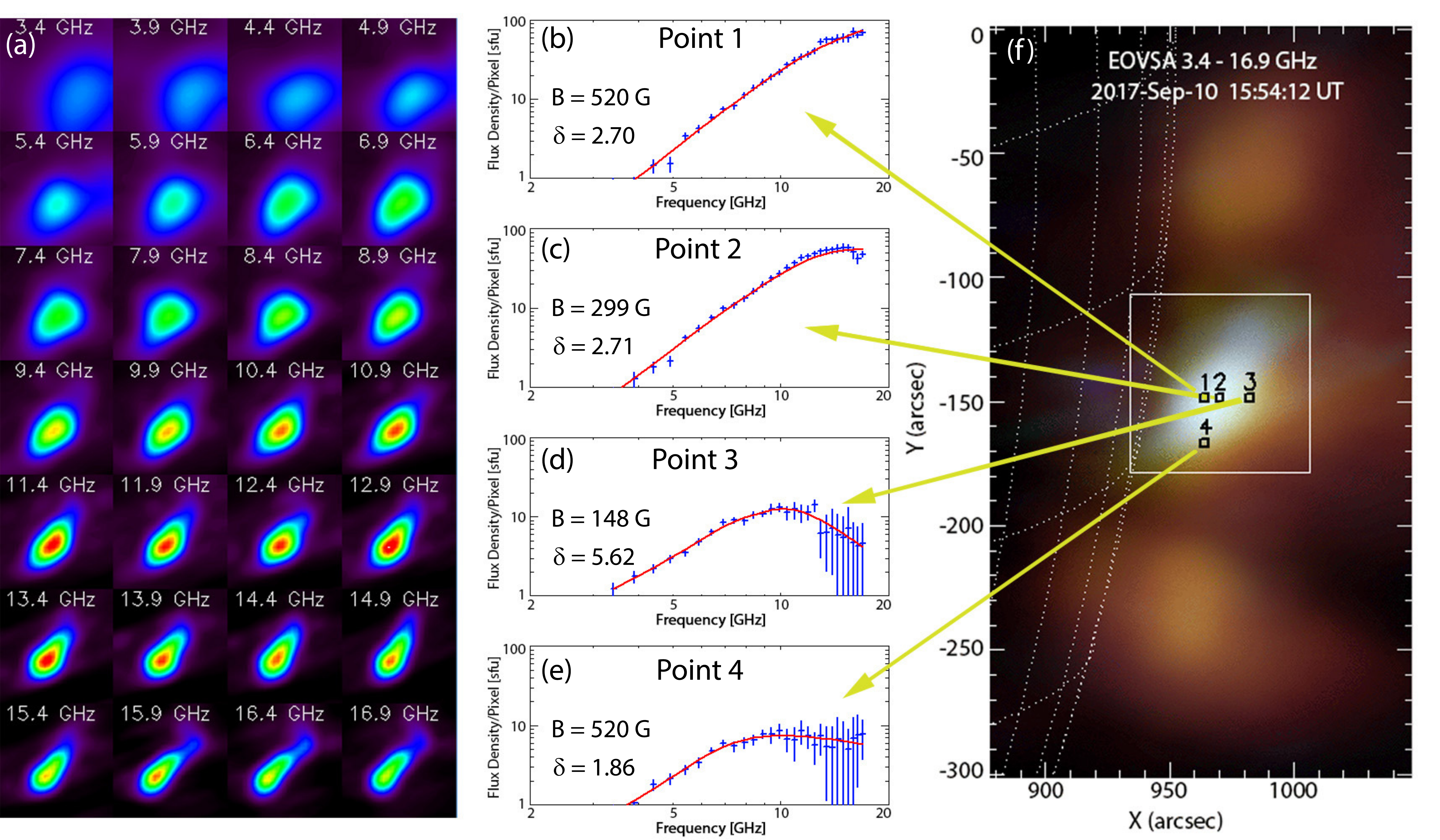}
\caption{\label{fig2}
A figure analogous to Fig.~\ref{fig1}, but this time with real data, from the event of 2017-Sep-10. (a) Individual images at 28 frequencies, from the location of the white box in the overview image in panel f. (b-e) Measured flux-density spectra (points with error bars) in single pixels of the images in panel a, corresponding to locations 1-4 marked in panel f, and corresponding multi-parameter fits (red lines). (f) A ``true-color'' representation of the EOVSA data cube, combining images at the 28 frequencies shown in panel a. (Adapted from \citealt{2018ApJ...863...83G}.)
 }
\end{figure}

In this example we focus on a single time, around 15:54:10 UT, shown as a ``true-color" image in Figure~\ref{fig2}f, where images at 28 frequencies are apportioned different red-green-blue weights according to their frequency.  This time is during the initial, rapid rise of a flux rope seen in EUV images of the event.  EOVSA shows strong radio emission below the inferred X-type reconnection point, with radio centroids located at steadily lower heights as frequency increases, shown by the shading of the central source from red to bluish-white. The EOVSA source shape also changes with frequency, from a cusp-shaped source at mid-frequencies (e.g. the 7.9~GHz image in Fig.~\ref{fig2}a), evolving toward a more loop-like shape at higher frequencies (15.9~GHz image).  At frequencies below about 5 GHz, two outlier sources appear (Fig.~\ref{fig2}f) that are associated with the legs of a larger loop.

These position and morphology dependences on frequency correspond to position-dependent spectral shapes, in analogy to our simulation described above.  This is demonstrated in Figure 2b-e, which are spectra measured at locations 1-4 marked in Figure~\ref{fig2}f. The points 1-3 located at different heights along the axis, and point 4 off to one side. The magnetic field strength inferred for each location is again indicated.

\begin{figure}\centering
\includegraphics[width=1.0\textwidth, clip]{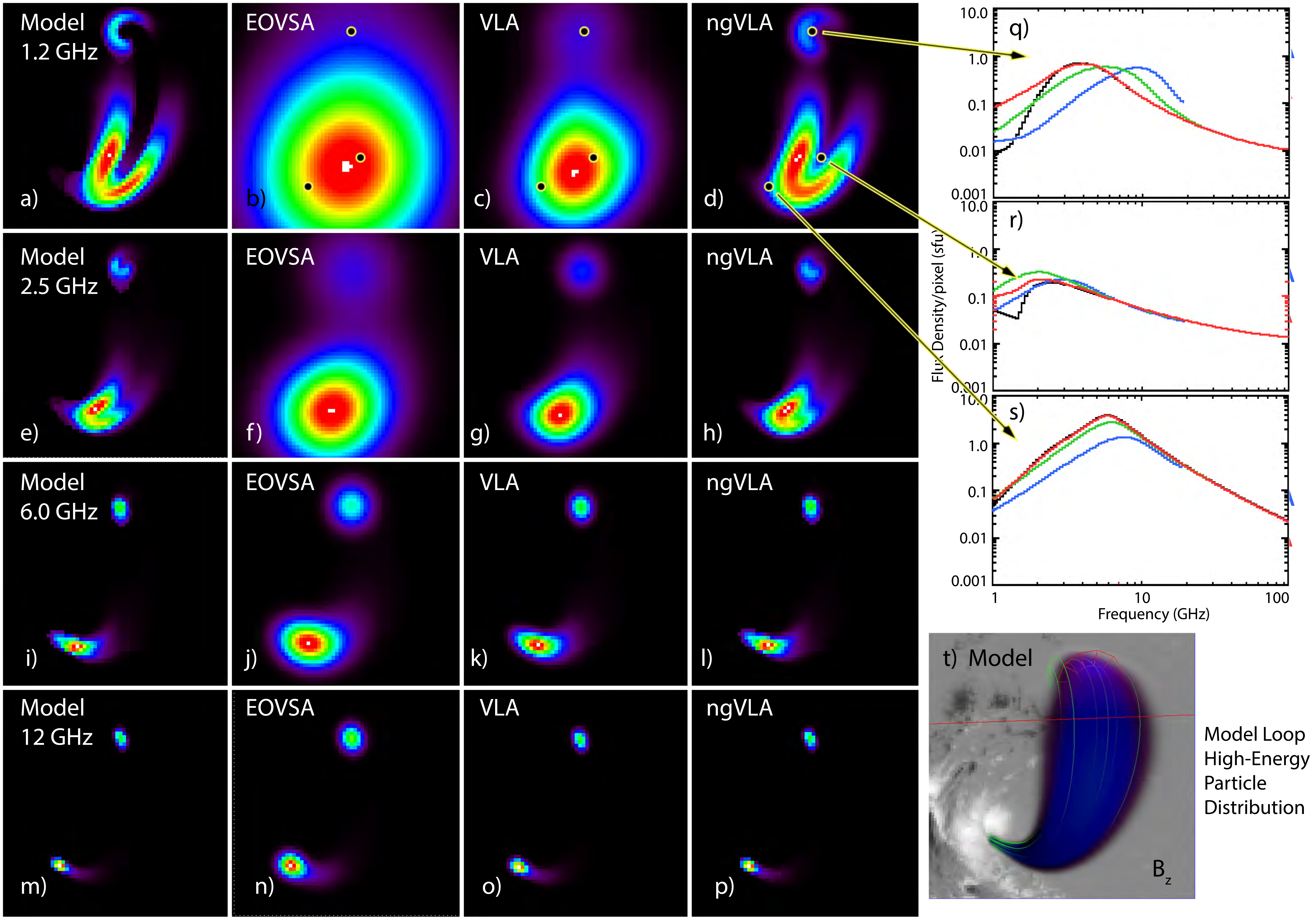}
\caption{\label{fig3}
A simulation of the flaring loop associated with the 2017 September 04 flare, as observed by EOVSA, and its appearance as observed by the VLA and ngVLA. The left column (panels a, e, i, m) show the microwave emission calculated from the model. The next three columns shown the same emission as would be seen with the resolution of EOVSA, VLA, and ngVLA, respectively.  The plots in the right column (panels q, r, s) show spectra from three spatial locations shown by the arrows, derived from the simulation for each of the four cases, model (black), EOVSA (blue), VLA (green) and ngVLA (red). Panel (t) is a rendering of the nonthermal electron distribution from the model loop, superposed on a line-of-sight magnetogram. No coronal scattering is included.}
\end{figure}

The above example thus demonstrates that, with the availability of spatially resolved microwave spectra, the diagnostic information inherent in GS emission can be exploited to provide measurements of the dynamically changing parameters in the flaring region. The ngVLA can improve on current capabilities in a few fundamental aspects. First, the ngVLA's combination of long baselines and excellent filling of smaller spacings will provide the uv coverage needed to extend imaging of solar flares to much higher spatial resolution than is currently possible.  The current VLA has configurations with long spacings, but the lack of shorter spacings makes the array essentially useless for solar work beyond the C configuration (with 3 km maximum spacing). This is illustrated in Figure~\ref{fig3}, from a simulation of the flaring loop associated with the 2017-Sep-04 flare, as observed by EOVSA.  The model used to calculate the radio emission is shown in panel (t) at the lower right, where the blue shading represents the volume distribution of high-energy electrons.  The first column (panels a, e, i, m) show the microwave emission at 1.2, 2.5, 6 and 12 GHz, respectively, as calculated from the model.  Note the bifurcation of the loop at low frequencies, where the loop edges are bright, but not its center.  This bifurcation is not captured in either EOVSA (1.2 km maximum spacing, nor the VLA C configuration (3 km maximum spacing). For the ngVLA images, spacings up to 12 km were assumed, although even longer spacings may be useful in some cases. When spectra at various points in the four data cubes are compared (panels q, r, and s), it is clear that higher resolution results in a more faithful measurement of the flux density/pixel, and features like the bifurcation of the loop become apparent. It should be noted that for the simulation in Figure~\ref{fig3} we have ignored scattering in the corona, which can be a limiting factor in the smallest angular size of flaring sources, as discussed further in \S~\ref{sec6}. An important strength of the ngVLA design is that uv limits can be expanded on a case-by-case basis as permitted by the inherent limits of the spatial structure in the observed source.

Second, by extending the spectral coverage towards much higher frequencies (compared with EOVSA), ngVLA will allow probing emission produced by relativistic charged particles with energies up to a few to a few dozens of MeV. This will permit quantification of spectral breaks and cutoffs and, thus, measure the highest energies to which nonthermal electrons can be accelerated. Third, having the high-frequency observations will open a new window of probing acceleration of ions. Indeed, the accelerated ions produce secondary relativistic positrons, which generate GS emission with the same intensity as electrons (provided those two populations have the same energy and angular distributions), but with the opposite circular polarization \citep{Fl_etal_2013}. Thus, given the magnetic field direction at the GS source, we can distinguish the positron contribution from the electron one, and in this way obtain additional constraints on acceleration of the associated ions in the flares. Finally, the greatly improved dynamic range of the images will not only provide higher-quality spectra for fitting, but will also reveal highly important, but generally much weaker, primary acceleration sources in the presence of the strong, trapped component that dominates after the first minutes of a flare.

\section{Synergies with Hard X-rays and Other Data}

In addition to the radio emissions already mentioned, highly energetic electrons also emit bremsstrahlung as they undergo collisions with ambient ions; this radiation shows up prominently in the HXR range (above a few keV).  Bremsstrahlung is highly weighted by plasma density and so appears most strongly at the dense chromospheric footpoints of flaring loops, but coronal emission at or near acceleration sites can also be observed.  Radio and HXR observations are the \textit{only} direct diagnostics of accelerated electrons in the solar atmosphere. Since they sometimes probe different energy regimes and are sensitive to different ambient parameters (e.g. magnetic field or plasma density), they naturally complement each other in tracing and diagnosing the flare-accelerated electrons \citep[e.g.,][and references therein]{Fl_etal_2016_narrow,2011SSRv..159..225W}.

Figure 2 has already shown an example of combining observations of coherent radio bursts and HXR data to trace accelerated electrons at or near the flare energy release site. Another example is given by \citet{bain2012}, who used the Nancay Radioheliograph to image a type II radio burst (another type of coherent radio emission) associated with a CME shock, while HXR images and spectra of the CME illuminated the accelerated electrons trapped within the ejection \citep{glesener2013}. More recently, \citet{musset2018} used measurements from RHESSI \citep{2002SoPh..210....3L} to investigate diffusive trapping of electrons from keV to tens of keV scale in flare loops, and compared these results with those found for more energetic (hundreds of keV), gyrosynchrotron-emitting electrons. The use of HXRs and gyrosynchrotron emission together with 3D modeling was found to be a powerful diagnostic of the flare-accelerated electrons \citep{Fl_etal_2016,2018ApJ...852...32K}. By studying the radio and HXR emissions together, a relatively full picture of the accelerated electrons was uncovered.

A variety of new HXR instrumentation is on the horizon or, in some cases, online already.  The \textit{Nuclear Spectroscopic Telescope Array} \citep{2013ApJ...770..103H}, an astrophysics mission in operation since 2012, is a direct-imaging HXR telescope that performs 2-4 solar observations per year, often in coordination with other solar observing campaigns \citep{grefenstette2016}.  The solar-dedicated Spectrometer/Telescope for Imaging X-rays (STIX) instrument will launch in the coming years as part of the \textit{Solar Orbiter (SO)} spacecraft \citep{krucker2016}.  STIX provides imaging and spectroscopy of HXRs from 4 to 150 keV with 1 keV spectral resolution.  In addition, the \textit{Focusing Optics X-ray Solar Imager (FOXSI)} is currently under consideration for launch in 2022 and would provide directly focused imaging spectroscopy of solar HXRs for the first time from a solar-dedicated spacecraft.  \textit{FOXSI} and ngVLA will be highly complementary.

\section{Requirements for the ngVLA}\label{sec6}

While the ngVLA is not optimized for observing solar flares, the basic attributes of the reference design are nevertheless compelling. It provides continuous frequency coverage from 1.2-50~GHz and 70-116~GHz, and contains a multitude of antennas on baselines less than a few km. The requirement for angular resolution at radio wavelengths for flare observations is $\theta\approx 20''/\nu_{GHz}$; i.e., $1''$ at a frequency of 20 GHz, or projected baselines out to $\approx\!3$~km. The reason that very high resolution is not needed, in general, is that ``coronal seeing'' imposes limits on the usable angular resolution \citep{1994ApJ...426..774B}. However, it is likely that coronal scattering varies widely on a case-by-case basis, and as noted earlier, the continuously filled uv-plane of the ngVLA will allow the maximum spatial resolution to be tailored for each case. Of paramount importance, too, is continuous frequency coverage of the bandwidth spanned by the coherent and incoherent emissions from a solar flare. Unfortunately, the ngVLA reference instrument does not provide frequency coverage below 1.2 GHz, although the idea of commensal or standalone arrays operating at low frequencies is being explored. However, the 1.2-3.5 GHz coverage will still be of great interest for coherent emissions associated with magnetic energy release. The frequency coverage provided by the higher frequency bands is for observations of the incoherent gyrosynchrotron and thermal free-free emission from solar flares. In many cases, coverage up to 50~GHz will be sufficient but studies of the most energetic flares will require coverage in 70-116~GHz band, too (although the small field of view in this band may be problematic).

Ideally, it will be possible to observe up to 6 ngVLA frequency bands simultaneously during solar flares, depending on the science objectives, dynamic range, and image fidelity requirements. The only practical way to achieve this is through the use of subarrays. With more than 120 antennas available within the required $3+$~km footprint, the array could be divided into as many as 5-6 independent subarrays, one for each frequency band, comprised of 20-25 antennas.

The time resolution required depends very much on the specific science objectives of a given observations but, loosely, observations of coherent bursts in the 1.2-3.5~GHz band require a time resolution of 10~ms or less. For observations of incoherent emissions a time resolution of 100~ms or more will be sufficient. Requirements for spectral resolution are modest: $\sim 0.1\%$ for coherent emissions and $\sim 1\%$ for incoherent emissions; i.e., more granular than needed for RFI excision.

It is important to recognize that the intense radio emissions observed on the Sun during a solar flare can introduce $10^7$-$10^9$ Jy to the system. Provisions must be made to observe and calibrate these signals without undue impact on ngVLA system electronics. This is not an insurmountable problem; the JVLA supports solar observations, as did the VLA before it, albeit in a non-optimum manner. If a signal attenuation and flux calibration strategy are designed in from the start, the ngVLA will be able to support a broad program of solar physics, including flares.

Finally, from an operational perspective, it should be noted that because flare occurrence and location cannot (yet) be reliably predicted, the ngVLA will need a flexible dynamic scheduling system for solar flare programs (similar to that currently supported by the JVLA), in order to maximize the probability of successful flare observations.

To summarize, solar flares continue to pose a number of fascinating challenges. Two have been been briefly discussed here - magnetic energy release and electron acceleration. More broadly, the ngVLA can make significant contributions to understanding the physics of flares if it supports dynamic imaging spectroscopy in all frequency bands.

\acknowledgements This work was supported by NSF grants AST-1615807, AST-1735405, AGS-1654382, AGS-1723436, AGS-1817277 and NASA grants NNX14AK66G, NNX16AL67G,  80NSSC18K0015, NNX17AB82G,
and 80NSSC18K0667.



\end{document}